\documentclass{emulateapj}
\usepackage[caption=false]{subfig}
\begin{document}
\title{Direct Exoplanet Detection with Binary Differential Imaging\footnotemark[*]}
\footnotetext[*]{This paper includes data obtained at the 6.5 m Magellan Telescopes located at Las Campanas Observatory, Chile.}
\author{Timothy J. Rodigas\altaffilmark{1,4}, Alycia Weinberger\altaffilmark{1}, Eric E. Mamajek\altaffilmark{2}, Jared R. Males\altaffilmark{3,5}, Laird M. Close\altaffilmark{3}, Katie Morzinski\altaffilmark{3}, Philip M. Hinz\altaffilmark{3}, Nathan Kaib\altaffilmark{1}}

\altaffiltext{1}{Department of Terrestrial Magnetism, Carnegie Institute of Washington, 5241 Broad Branch Road, NW, Washington, DC 20015, USA; email: trodigas@carnegiescience.edu}
%\altaffiltext{2}{NASA Goddard Space Flight Center, Exoplanets $\&$ Stellar Astrophysics Laboratory, Code 667, Green- belt, MD 20771}
%\altaffiltext{3}{Space Telescope Science Institute, Baltimore, MD 21218, USA}
\altaffiltext{2}{Department of Physics $\&$ Astronomy, University of Rochester, Rochester, NY 14627, USA}
\altaffiltext{3}{Steward Observatory, The University of Arizona, 933 N. Cherry Ave., Tucson, AZ 85721, USA}
%\altaffiltext{5}{INAF - Osservatorio Astrofisico di Arcetri, Largo E. Fermi 5, I-50125, Firenze, Italy}
\altaffiltext{4}{Hubble Fellow}
\altaffiltext{5}{NASA Sagan Fellow}

%\altaffiltext{7}{Hubble Fellow}
%\altaffiltext{4}{Department of Physics and Astronomy, University of Rochester, Rochester, NY 14627-0171, USA}
%\altaffiltext{5}{Rockhurst University, 1100 Rockhurst Rd, Kansas City, MO 64110, USA}
%\altaffiltext{6}{University of Toronto, 50 St George St., Toronto, ON M5S 1A1, Canada}
%\altaffiltext{7}{School of Earth and Space Exploration, Arizona State University, PO Box 871404, Tempe, AZ 85287-1404, USA}
%\altaffiltext{8}{Large Binocular Telescope Observatory, University of Arizona, Tucson, AZ 85721, USA}
%\altaffiltext{9}{University of Virginia, Department of Astronomy, 530 McCormick Road, Charlottesville, VA  22903, USA}

\newcommand{\about}{$\sim$~}
\newcommand{\mj}{M$_{J}$}
\newcommand{\degrees}{$^{\circ}$}
\newcommand{\arcseconds}{$^{\prime \prime}$}
\newcommand{\asec}{$\arcsec$}
\newcommand{\fasec}{$\farcs$}
\newcommand{\lprime}{$L^{\prime}$}
\newcommand{\ks}{$Ks$~}
\newcommand{\mjyasec}{mJy/arcsecond$^{2}$}
\newcommand{\microns}{$\mu$m}

%\shorttitle{something}
\shortauthors{Rodigas et al.}

\begin{abstract}
Binaries are typically excluded from direct imaging exoplanet surveys. However, the recent findings of Kepler and radial velocity programs show that planets can and do form in binary systems. Here, we suggest that visual binaries offer unique advantages for direct imaging. We show that Binary Differential Imaging (BDI), whereby two stars are imaged simultaneously at the same wavelength within the isoplanatic patch at high Strehl ratio, offers improved point spread function (PSF) subtraction that can result in increased sensitivity to planets close to each star. We demonstrate this by observing a young visual binary separated by 4\asec ~with MagAO/Clio-2 at 3.9 \microns, where the Strehl ratio is high, the isoplanatic patch is large, and giant planets are bright. Comparing BDI to angular differential imaging (ADI), we find that BDI's 5$\sigma$ contrast is \about 0.5 mags better than ADI's within \about 1\asec ~for the particular binary we observed. Because planets typically reside close to their host stars, BDI is a promising technique for discovering exoplanets in stellar systems that are often ignored. BDI is also 2-4$\times$ more efficient than ADI and classical reference PSF subtraction, since planets can be detected around both the target and PSF reference simultaneously. We are currently exploiting this technique in a new MagAO survey for giant planets in 140 young nearby visual binaries. BDI on a space-based telescope would not be limited by isoplanatism effects and would therefore be an even more powerful tool for imaging and discovering planets.
\end{abstract}

\section{Introduction}
A fundamental limit to directly imaging planets on solar system scales is the ``speckle floor." This is the region extending out to \about 1\fasec 5 from the star that is dominated by speckles: time-varying aberrations in the point spread function (PSF). Because the locations and brightnesses of the speckles change slowly with time, it is difficult to fully subtract them, even with the help of adaptive optics (AO). Speckle noise thus dominates over photon and sky noise close to the star \citep{speckles}, limiting our ability to directly detect faint planets at small separations.

Angular differential imaging (ADI; \citealt{adi}), in which the target star serves as its own PSF reference, has helped improve contrast close to the star, but only to a certain level and usually at the cost of attenuating the fluxes of the off-axis companions. ADI also requires a sufficient amount of sky rotation to occur throughout the observations: the less the sky rotates, the more a companion's flux at a given separation is attenuated. This means that planets located in the speckle floor region will be more attenuated than those farther out, making their detection even more challenging. 

Simultaneous differential imaging (SDI; \citealt{sdi,speckles}) can in theory completely remove speckles because the reference PSF is the target itself imaged simultaneously at different wavelengths, but it requires the faint companions to have unique and prominent spectral features over narrow wavelength regions compared to the stars--an assumption that is so far uncertain. Furthermore, the different Strehl ratios of the two simultaneously-acquired images preclude true photon noise limited PSF subtraction. For these reasons, it has been difficult for current and past direct imaging surveys to detect low-mass planets at small separations. 

At wider separations, a handful of planets have been directly imaged (e.g., HR 8799 bcde, \citealt{marois,hr87994thplanet}; Beta Pic b, \citealt{betapicoriginal,betapic}; HD 95086 b, \citealt{95086confirm}). However, with such a small sample size, it is difficult to make statistical inferences on the general properties of giant planets, and even harder to constrain formation and evolution theories. To alleviate this problem, the number of imaged exoplanets must increase.

We know from several imaging surveys that giant planets are rarely found at wide separations \citep{nielsennici,wahhaj,jansonsurvey,billermovinggroups,maireplanetsurvey}. We also know from radial velocity (RV) and transit surveys that giant planets \textit{do reside} at small separations (semimajor axis $a < 5$ AU; \citealt{butler,cumming,keplerstats}). Therefore future imaging surveys must improve detection sensitivity \textit{close} to the star.

Detecting planets at close separations--without building new telescopes/instrumentation--calls for a different PSF subtraction technique. To avoid self-subtraction (which plagues ADI), the target star cannot be used as its own PSF reference. To avoid atmospheric requirements in the companion (which plague SDI), the target and PSF reference must be imaged at the same wavelength. ``Classical" PSF subtraction, wherein observations of a reference PSF are interleaved with observations of the target, is inefficient (two or more stars imaged to search for planets around one) and suffers from a time-varying PSF. Therefore the only remaining option is to observe \textit{two (or more) stars at the same time at the same wavelength}. 

This is not a new idea. Speckle holography has been demonstrated as a viable technique to infer the presence of faint companions using the PSFs of multiple simultaneously imaged stars \citep{holography,holography2,koreskobinaries}, as long as they are within the isoplanatic patch at the observed wavelength \citep{isoplanatic}. In space, the Hubble Space Telescope has been used to search for circumstellar dust in a binary by using each component's PSF \citep{hstbdi}. From the ground with AO, \cite{mcelwainthesis} used Keck/OSIRIS to search for companions in several close visual binaries (no companions were found).

We can improve on these previous studies in three ways. First, we can take advantage of modern AO systems, which now offer high Strehl ratio imaging, especially at long wavelengths. This improves our ability to detect faint planets close to their host stars. Second, we can image near 4 \microns. At this wavelength, gas giant planets are thought to be bright \citep{burrows,baraffe}, and the isoplanatic patch is large (\about 10-30\asec). A larger isoplanatic patch means binaries with larger projected separations will have nearly identical PSFs. Indeed, \cite{heinzesurvey} imaged several binaries at 4-5 \microns ~with the MMT deformable secondary AO system and showed that subtracting each PSF from the other could be more effective than ADI. This is possible because the target is no longer used as its own PSF reference, so the flux from the companion will \textit{never} be attenuated by self-subtraction. Furthermore because the PSF is imaged at the same wavelength as the target, the companion is not required to have unique spectral features relative to the star (as with SDI). 

Finally, we can take advantage of advanced post-processing algorithms like principal component analysis (PCA, \citealt{pca}), which has been shown to significantly improve PSF subtraction. This raises the possibility of completely removing the speckle floor from each star during the data reduction stage and thus reaching the photon noise limit. The photon noise limit is a critical threshold because, once reached, longer integrations result in increased sensitivity. This is not the case with ADI, where at small separations speckle noise dominates and adequate sky rotation, rather than integration, is preferred. 

It is now clear that planets can and do form in binaries \citep{keplervisualbinary,friendsofhotjupiters2}. With the added benefit of potentially superior PSF subtraction, visual binaries should no longer be excluded from direct imaging planet surveys. In this paper, we show that Binary Differential Imaging (BDI), in which two stars are imaged simultaneously, at high Strehl ratio, at the same wavelength, and within the isoplanatic patch, both improves detection sensitivity at small separations by \about half a magnitude and is 2-4$\times$ more observationally efficient than current ground- and space-based direct imaging surveys, respectively. In Section 2, we describe the BDI technique. In Section 3, we present proof of concept data obtained at the Magellan Clay telescope using MagAO \citep{magao} and the near-infrared (NIR) camera Clio-2 \citep{suresh}. In Section 4 we compare the achieved contrast of ADI and BDI. In Section 5 we discuss the technical and scientific benefits of BDI, on both ground- and space-based telescopes. In Section 6 we summarize and conclude.

\section{Binary Differential Imaging: Expected Signal-to-Noise}
\label{sec:theory}
Two stars are simultaneously imaged on a detector at the same wavelength with AO. The separation of the stars is within the isoplanatic patch at that wavelength. Star A is brighter than Star B by $f = 10^\frac{M_A-M_B}{-2.5}$, where $M$ is the magnitude of a given star at the chosen wavelength. Consider two pixels (Pixel A and Pixel B) each having the same coordinates (radius, position angle) relative to each star. Assume that a hypothetical planet resides around Star A and some of its flux falls in Pixel A; Star B has no planets. We will calculate the total signal to noise (S/N) of the planet's flux that results from subtracting one pixel from the other. In real high-contrast imaging PSF subtraction, many pixels comprise an image that is subtracted from another image, but the case is more simply explained using a single pixel.

The total flux in Pixel A is given by
\begin{equation}
T_{A} = I_A + I_P + S,
\end{equation}
where $I_A$ is the flux from Star A, $I_P$ is the flux from the planet, and $S$ is the flux from the sky. The sky flux is removed by subtracting blank sky ($S_2$) so that $T_A \rightarrow I_A + I_P + S - S_2$. We will assume that the sky flux is completely removed so that the total flux in Pixel A is just the flux from the star and planet. The same is true for Pixel B (without planet flux). In both cases, however, both read noise ($R$) and photon noise from the star ($\sigma_{I}$) and sky ($\sigma_{sky}$) contribute to the uncertainty. We will assume that the photon noise from the planet ($\sigma_{I_P}$) is negligible. Therefore the total noise in Pixel A is given by:
\begin{eqnarray}
\sigma_{T_A} & = & \sqrt{ \frac{\partial T_A}{\partial I_A}^{2} \sigma_{I_A}^2 + \frac{\partial T_A}{\partial S}^{2} \sigma_{S}^2 +  \frac{\partial T_A}{\partial S_2}^{2} \sigma_{S_2}^2 + 2R^2} \nonumber \\ 
 & = & \sqrt{\sigma_{I_A}^2 + 2\sigma_{sky}^2 + 2R^2} \nonumber \\ 
 & = & \sqrt{I_A + 2S + 2R^2},
\end{eqnarray}
where the stellar photon noise $\sigma_{I_A} = \sqrt{I_A}$, and we have assumed that $\sigma_S = \sigma_{S_{2}} = \sigma_{sky} = \sqrt{S}$ and that the read noise is the same in the target and sky images. The total noise in Pixel B is similarly given by
\begin{eqnarray}
\sigma_{T_B} & = & \sqrt{I_B + 2S + 2R^2} \nonumber \\
  & = & \sqrt{I_A/f + 2S + 2R^2},
\end{eqnarray}
where we have made the substitution $I_B = I_A/f$. 

Scaling and subtracting the Pixel B flux from Pixel A, we then have:
\begin{equation}
\label{eqn:pixelA}
F = T_{A} - f T_B,
\end{equation}
which = $I_P$ if the PSFs are identical. However, this is an ideal case and is unrealistic. Instead, there will probably be some slight differences in the PSFs due to non-isoplanatism. Therefore the remaining flux will be $F = I_P + i$, where $i$ is this (positive or negative) extra flux. We can define a new variable $G = F - i = T_A - f T_B - i$ and calculate the total noise:
\begin{eqnarray}
\label{eqn:residual}
\sigma_{BDI} & = & \sqrt{ \frac{\partial G}{\partial T_A}^{2} \sigma_{T_A}^2 + \frac{\partial G}{\partial T_B}^{2} \sigma_{T_B}^2 + \frac{\partial G}{\partial i}^2 \sigma_{i}^2} \nonumber \\
 & = & \sqrt{I_A + 2S + 2R^2 + f^{2} (I_A/f + 2S + 2R^2) + \sigma_i^2} \nonumber \\
 & = & \sqrt{(1 + f) I_{A} + (1 + f^{2}) (2S + 2R^2) + \sigma_i^2} \label{eqn:sigmaBDI},
\end{eqnarray}
where we have assumed that the uncertainty in the scaling factor $\sigma_f \approx 0$. In reality, this uncertainty will be non-zero, but it is typically much smaller than the other sources of noise and is therefore ignored here. 

The total S/N of the planet in the BDI case is given by
\begin{equation}
\label{eqn:snbdi}
S/N_{BDI} = \frac{c ~ I_A}{\sqrt{(1 + f) I_{A} + (1 + f^{2}) (2S + 2R^2) + \sigma_i^2}}.
\end{equation}
Here we have made the substitution $I_P = c ~ I_A$, where $c$ is the contrast of the planet relative to the host star at the location of the planet.

It is thus evident that as $f$ increases beyond the ideal $f = 1$, $S/N_{BDI}$ decreases. In the case of Pixel B minus Pixel A, we again have Eq. \ref{eqn:snbdi} except $I_A \rightarrow I_B$ and $f \rightarrow 1/f$. In this case, $S/N_{BDI}$ is maximized as $f \rightarrow \infty$. If the two PSFs are truly identical, then $\sigma_i = 0$ and the noise in BDI PSF subtraction is limited by the brightness ratio of the two imaged stars, with better PSF subtraction (in terms of noise) occurring when the brighter star PSF is subtracted from the fainter star PSF. If the PSFs are not identical, then the noise is limited by the amount of non-isoplanatism ($\sigma_i$).

%We can also see that as $S, R \rightarrow 0$, $\sigma_{BDI}$ becomes limited by stellar photon noise.

\subsection{Comparison to ADI}
In the case of ADI PSF subtraction, Pixel B and Pixel A both correspond to Star A, but Pixel B is recorded on the detector some time after Pixel A. We will again assume that Pixel A contains the planet flux, and that Pixel B contains no planet flux. However, to account for the self-subtraction that is inherent to ADI, we will assume that $I_P \rightarrow a(r) I_P$, where $a$ is the attenuation factor that depends on the distance from the star $r$. 

In theory, the brightness ratio of Pixel A and B ($f$) is unity, but the PSFs will \textit{never} be identical. Therefore the noise is limited by the differences in the PSFs. To illustrate, we can derive $\sigma_{ADI}$ in the case where we have the same star imaged at two different times resulting in two different total fluxes, $T_{A}$ (which contains the planet flux) and $T_{B}$. The subtraction of the two pixels gives $F = T_{A} - T_{B} = a(r) I_P + k$, where $k$ is the residual flux that results from the non-identical PSF subtraction. As before we can define a new variable $H = F - k = T_{A} - T_{B} - k$ and calculate the noise:
\begin{eqnarray}
\label{eqn:residualADI}
\sigma_{ADI} & = & \sqrt{ \frac{\partial H}{\partial T_{A}}^2 \sigma_{T_{A}}^2 + \frac{\partial H}{\partial T_{B}}^2 \sigma_{T_{B}}^2 + \frac{\partial H}{\partial k}^2 \sigma_{k}^2} \nonumber \\
 & = & \sqrt{2 I_{A} + 2 (2S + 2R^2) + \sigma_k^2} \label{eqn:sigmaADI}. 
\end{eqnarray}
The total S/N of the planet in the ADI case is given by
\begin{equation}
\label{eqn:snadi}
S/N_{ADI} = \frac{a(r) ~ c ~ I_{A}}{\sqrt{2 I_{A} + 2 (2S + 2R^2) + \sigma_k^2}}.
\end{equation}

\begin{figure}[t]
\centering
\includegraphics[width=0.49\textwidth]{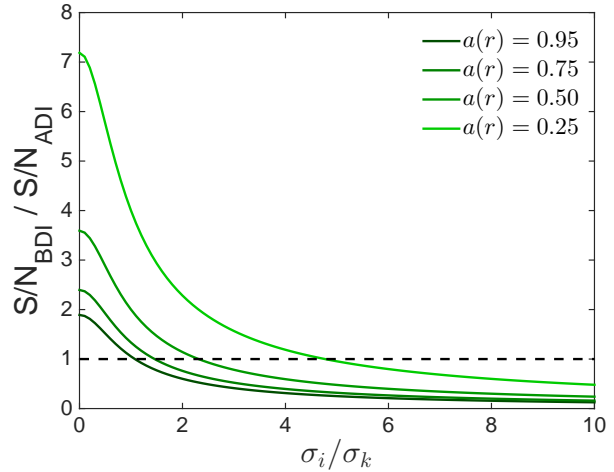}
%\subfloat[]{\label{fig:contrastcurve}\includegraphics[width=0.49\textwidth]{contrastcurve.eps}}
%\includegraphics[width=0.49\textwidth]{BDIpercentA.eps}
\caption{Ratio of expected BDI S/N (Eq. \ref{eqn:snbdi}) to expected ADI S/N (Eq. \ref{eqn:snadi}) as a function of $\sigma_i/\sigma_k$, where we have set $R = 1$, $S = 50$, $I_A = 100$, $\sigma_k = 30$, and $c = 10^{-3}$. The S/N ratio also depends on the attenuation factor $a(r)$, which affects only ADI. The dashed line denotes the division between BDI or ADI being preferred. As $a(r)$ decreases, BDI is preferred even at large $\sigma_i$ values. When $a(r)$ is close to 1, BDI will only win out if $\sigma_i$ is close to $\sigma_k$. This leads to the expectation that in real data, BDI should supersede ADI close to the star.}
\label{fig:snplots}
\end{figure}

Eq. \ref{eqn:sigmaADI} is identical to the ideal ($f = 1$) BDI noise (Eq. \ref{eqn:sigmaBDI}), except for the different noise term $\sigma_k$ (compared to $\sigma_i$). This term (speckle noise) can in practice be estimated by calculating the standard deviation of pixels in an annulus at a given radius; it typically increases closer to the star and can be orders of magnitude larger than the other noise sources \citep{speckles}. Thus ADI will yield lower noise than BDI if $\sigma_k < \sigma_i$. However, $S/N_{ADI}$ is also limited by the attenuation factor $a(r)$. Fig. \ref{fig:snplots} shows the theoretical S/N for BDI (ideal, $f = 1$) and ADI, for different values of $\sigma_k$, $\sigma_i$, and $a(r)$. It will be our goal in this paper to determine with real data how these parameters affect the achieved S/N of each imaging technique.

\subsection{Comparison to Classical PSF Subtraction}
We can also compare BDI PSF subtraction to classical reference star PSF subtraction (sometimes referred to as ``Reference Differential Imaging," or RDI), which is typically employed for space-based telescopes. We will assume we are observing with a coronagraph on the the James Webb Space Telescope (JWST) NIRCam instrument. We will not compare to PSF ``roll" subtraction  \citep{hstroll} because JWST's limited roll angle (\about 10\degrees) will likely preclude exoplanet imaging using this technique, at least for close-in planets. 

The noise due to BDI PSF subtraction with NIRCam (no coronagraph\footnote{Coronagraphs (in the image plane) cannot be used on NIRCam with BDI because they change the shape of one PSF but not the other(s) in the FOV.}) is again given by Eq. \ref{eqn:sigmaBDI}. For simplicity, we will assume the ideal case of two identical binary stars ($f = 1$), no sky or read noise, and no noise due to the instrument (i.e., distortion), so that $\sigma_{BDI} = \sqrt{2 I_A}$. We will also assume $\sigma_i = 0$, since there are no isoplanatism effects in space. The signal in the BDI case is simply $c I_A$, so that:
\begin{equation}
\label{eqn:bdispace}
S/N_{BDI,space} = \frac{c ~ I_{A}}{\sqrt{2 I_{A}}}.
\end{equation}

The noise due to coronagraphic PSF subtraction is similar to the BDI case (again assuming an ideal $f = 1$), except for the extra noise due to PSF subtraction residuals, $\sigma_k$. Therefore we have: $\sigma_{coron} = \sqrt{2 g I_A + g^2 \sigma_k^2}$, where $g$ is the fractional decrease in flux that is caused by the coronagraph. The signal of the planet is $t c I_A$, where $t$ is the throughput of the coronagraph. The total S/N is then:
\begin{equation}
\label{eqn:coronspace}
S/N_{coron,space} = \frac{t ~ c ~ I_{A}}{\sqrt{2 g I_{A} + g^2 \sigma_k^2}}.
\end{equation}

Comparing the S/N terms, we can see that $S/N_{BDI,space} > S/N_{coron,space}$ when $\sigma_k > g^{-1} \sqrt{2 I_A (t^2 - g)}$. This relation can also be expressed as $\sigma_k > g^{-1} \sqrt{2 (t^2 - g)} \sigma_{I_A}$, where $\sigma_{I_A}$ is the stellar photon noise. For $g \approx 10^{-1}$ and $t = 0.81$ \citep{nircam}, the S/N from BDI PSF subtraction is larger if $\sigma_k \gtrsim 10 \sigma_{I_A}$. It is not yet clear what $\sigma_k$ will be for JWST/NIRCam, but this threshold is a useful diagnostic. Even if $\sigma_k$ is small (a few $\times$ the stellar photon noise), BDI may still be preferred because it can reach better IWAs than coronagraphic imaging at a minor cost of slightly larger noise. Furthermore, because BDI is essentially photon noise limited, with enough integration it can always supersede classical PSF subtraction.  

%While $\sigma_k$ is still unknown because JWST has not yet been launched, it is not unreasonable to expect $\sigma_k > $ \about few $\sigma_{I_A}$ at all separations. Therefore BDI should yield lower noise than classical space-based coronagraphic PSF subtraction. 

%Taking into account the fact that BDI preserves companion signal (unlike ADI) and can potentially reach lower noise levels than both ADI and classical space-based PSF subtraction, BDI should be considered a for PSF subtraction.

%We have shown above that the noise from BDI PSF subtraction should be smaller than in the ADI and space-based/classical PSF subtraction techniques. When we also take into account the \textit{signal} of a recovered companion, BDI should be the clear preferred method. Both BDI and classical PSF subtraction generally preserve the flux of the companion, unlike ADI. Because the noise will 

\section{Testing BDI: Observations and Data Reduction}
%\subsection{Observations and Data Reduction}
To test the feasibility of BDI--in particular, the similarity of two PSFs separated by several arcseconds on the sky--we carried out proof of concept observations of a binary with MagAO/Clio-2. We observed the young visual binary HD 37551, which consists of a G7 (primary) and K1 \citep{sacy} separated by \about 4\asec at a distance of 77$\pm$9 pc \citep{updatedhip}. The primary (henceforth Star A) has a V magnitude = 9.650$\pm$0.017 and the secondary (henceforth Star B) has V = 10.538$\pm$0.017 \citep{hd37551vmag}. HD 37551 is thought to be a member of the AB Dor moving group \citep{sacy,hd37551sptype}. However, the Li depletion pattern of the low-mass AB Dor members and their color-magnitude diagram positions appear to be indistinguishable with the Pleiades \citep{abdorage,abdorchem}, whose Li-depletion boundary age is 130$\pm$20 Myr \citep{pleiadesage}. We therefore adopt this age for HD 37551.

We observed the binary on the night of UT 26 November 2014 at the Magellan Clay telescope of the Las Campanas Observatory (LCO) in Chile. We obtained diffraction-limited AO images (300 modes) of the binary in the narrowband 3.9 \microns ~filter ($\Delta \lambda = 0.23$ \microns, $\lambda_c = 3.95$ \microns; hereafter [3.9]) using the narrow camera (plate scale = 0\fasec 01585 per pixel\footnote{\url{http://zero.as.arizona.edu/groups/clio2usermanual/}}). Star A served as the guide star for the AO system. The instrument rotator was turned off to enable ADI; therefore Star B rotated with the sky around Star A. Observing conditions were good, with the seeing \about 0\fasec 75. Every few minutes, we nodded the binary off the detector to acquire images of blank sky. Neither star's PSF was saturated in any of the recorded images.

After discarding frames of poor quality (e.g., open loop), we acquired 189 images of the binary for a total of 31.5 minutes of integration. The parallactic angle (PA) changed by 19.37\degrees ~throughout the observations. While obviously not a large amount of sky rotation, this is sufficient for directly comparing the detection sensitivities of ADI and BDI.

\begin{figure*}[t]
\centering
\subfloat[]{\label{fig:PSFA}\includegraphics[width=0.49\textwidth]{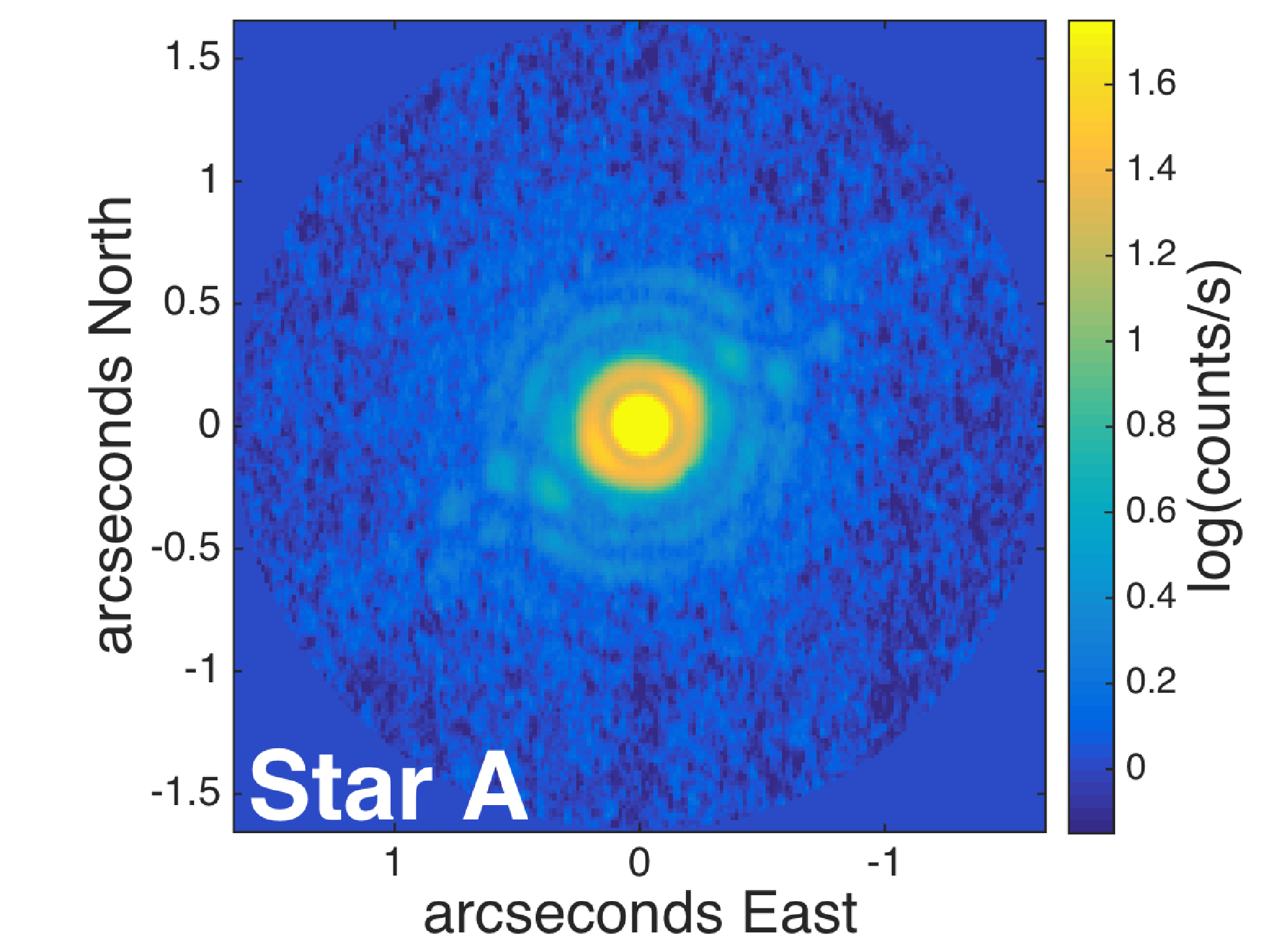}} 
\subfloat[]{\label{fig:PSFB}\includegraphics[width=0.49\textwidth]{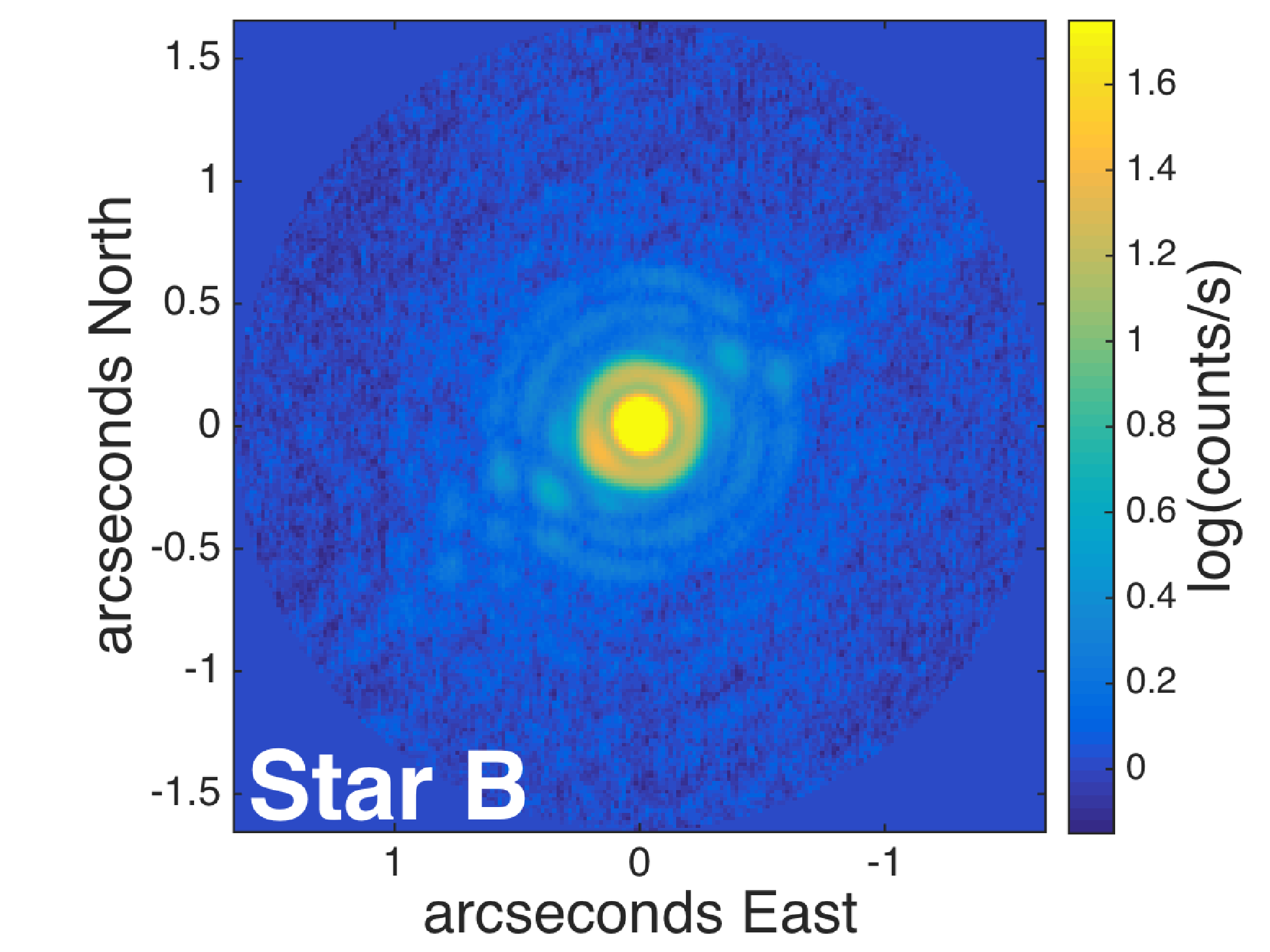}}
\caption{The HD 37551 binary imaged at [3.9]. (a). Median-combination of all the Star A images. (b) The same, but for the Star B images. Despite the brightness difference between the two stars (measured to be 0.86 mags at [3.9]) and the 4\asec ~separation, the PSFs are nearly identical. However, some small differences are evident that make the BDI PSF subtraction noise ($\sigma_i$) non-zero. The sky background residuals are also different, most likely due to variations in the detector arising from Star B rotating around Star A during the observations.}
\label{fig:PSFS}
\end{figure*}

All data reduction was performed using custom scripts in Matlab. We divided the images by the number of coadds, corrected them for non-linearity, and divided them by the integration time to obtain units of detector counts/s. To remove the sky background, for each image of the binary we constructed an optimal sky image using all the sky images, such that the noise in a small rectangular box far from the stars was minimized. Next we determined the sub-pixel location of each star by calculating the center of light inside a 0\fasec 25 aperture centered on the stars. We then registered and cropped the images of each star. Finally, we corrected for bad pixels. Fig. \ref{fig:PSFS} shows the median PSFs for Star A and Star B, respectively. As expected, the PSFs (and speckles) are nearly identical, despite the different stellar brightnesses and 4\asec ~separation.

Next, we performed ADI and BDI PSF subtraction. To simplify the comparison of the two techniques, we performed PSF subtraction on only Star A. Note that according to Section \ref{sec:theory}, the PSF subtraction noise for BDI will be higher in this case, so our BDI results should be viewed as conservative.

\subsection{ADI Reduction}
We reduced the Star A data using our custom ADI+PCA pipeline. Because the PSF remained in approximately the same location on the detector (aside from the off-chip sky nods), the PSF was very stable. This meant that fewer PCA modes were needed to achieve excellent PSF subtraction. Using fewer modes also helped to minimize companion attenuation, which was important here given the small amount of sky rotation ($< 20$\degrees). We settled on using 2 modes (out of a possible 189) to construct the PSF for each image. No rotation requirements were applied (i.e., all target images were available for PSF construction), because doing so does not necessarily increase recovered companion S/N \citep{tiffany2}.\footnote{See Section \ref{sec:limitations} for a justification of this choice.} After PSF subtraction, we rotated the images by their PAs to North-up, East-left and median-combined them to make the final image. No bright off-axis point-sources were detected in the final image. 

\subsection{BDI Reduction}
The BDI+PCA data reduction is in practice quite similar to the ADI+PCA reduction, with one difference: instead of the target itself serving as the PSF reference for PCA, the Star B images are used. As with ADI, the number of PCA modes was again a free parameter. The number of modes for BDI should not a priori be identical to the number of modes used in the ADI reduction, since the PSFs used by ADI and BDI are not necessarily identical. Furthermore, because ADI is already hampered by companion self-subtraction, fewer modes will be preferred for ADI to maximize recovered companion S/N. For BDI, we found that while increasing the number of modes did increase the signal-to-noise (S/N) of recovered artificial companions, the increase was marginal and at the expense of computation time. Therefore we settled on 10 (out of 189) modes, which was a good compromise between recovered S/N and efficiency. We rotated the PSF-subtracted images by their PAs to obtain North-up, East-left and median-combined them into a final image, which was additionally unsharp masked to remove residual background flux. No bright off-axis companions were detected in the final BDI image, in agreement with ADI.

%However, before feeding the images into the PCA routine, one also has the option of subtracting the mean Star A image from all the Star A images and the mean Star B image from all the Star B images. Doing this significantly lowers the noise in the final image, but it also partially attenuates the flux of any companions, especially at close separations (see Fig. \ref{fig:BDImean}). If the mean images are \textit{not} subtracted, the noise due to PSF residuals is higher, but the companion signal is preserved. Therefore the S/Ns of injected artificial companions are higher (Fig. \ref{fig:BDImean}). We adopted this latter (no mean subtraction) method for our nominal BDI data reduction.

%\begin{figure*}[t]
%\centering
%\subfloat[]{\label{fig:adi}\includegraphics[width=0.33\textwidth]{ADInormal.eps}} 
%\subfloat[]{\label{fig:bdiwmean}\includegraphics[width=0.33\textwidth]{BDImeansub.eps}}
%\subfloat[]{\label{fig:bdiwomean}\includegraphics[width=0.33\textwidth]{BDInormal.eps}}
%\caption{Examples of artificial companions injected and recovered by ADI (a), BDI with mean subtraction (b), and BDI without mean subtraction (c). While the first two methods yield better overall PSF subtraction (lower noise due to residuals), the companion flux is attenuated by self-subtraction, decreasing the companion S/N compared to BDI.}
%\label{fig:BDImean}
%\end{figure*}

\section{ADI vs. BDI: Comparing Contrast Limits}
\label{sec:fake}
%Fig. \ref{fig:BDImean} shows that ADI achieves better PSF subtraction than BDI, in terms of PSF residuals. This tells us that the binary PSFs are \textit{not} identical and are more dissimilar than the PSFs of a star imaged \about a few minutes apart (the ADI case). In other words, referring back to Section \ref{sec:theory}, $\sigma_i > \sigma_k$. Nonetheless, because BDI does not attenuate companion signal, it still has the potential to yield higher S/N detections of planets than ADI. 

\begin{figure*}[t]
\centering
\subfloat[]{\label{fig:ADIfakefar}\includegraphics[width=0.49\textwidth]{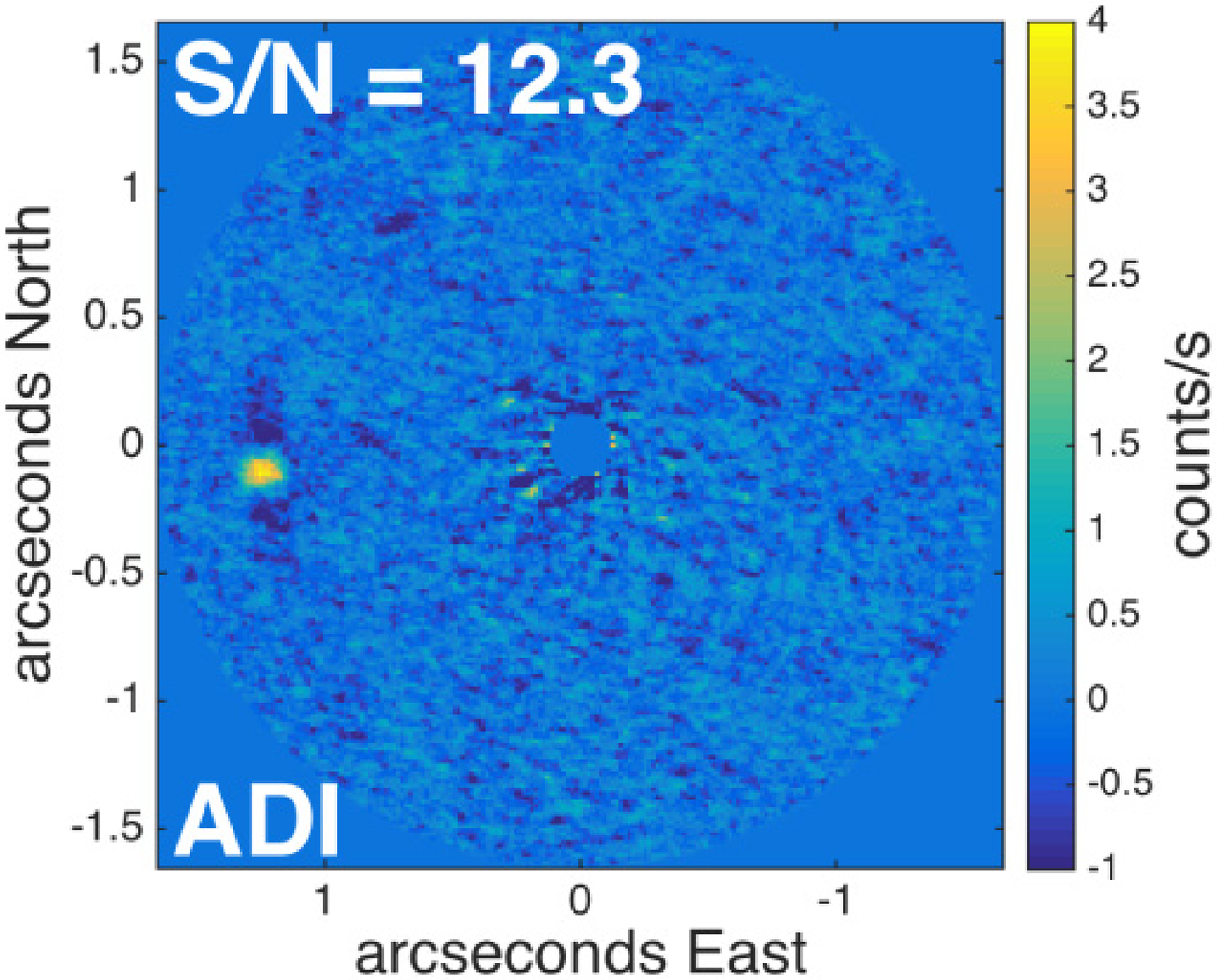}} 
\subfloat[]{\label{fig:BDIfakefar}\includegraphics[width=0.49\textwidth]{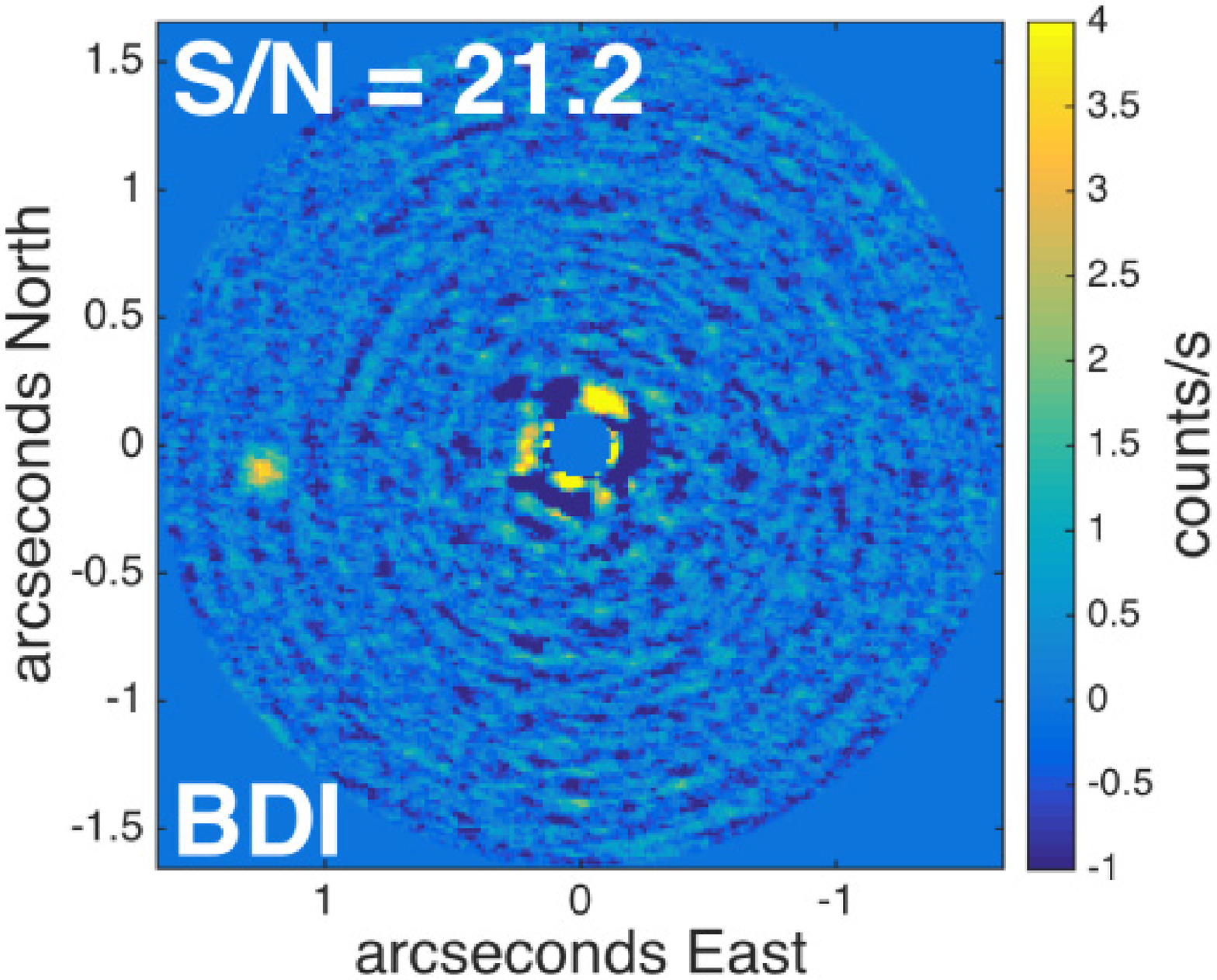}} \\
\subfloat[]{\label{fig:ADIfakeclose}\includegraphics[width=0.49\textwidth]{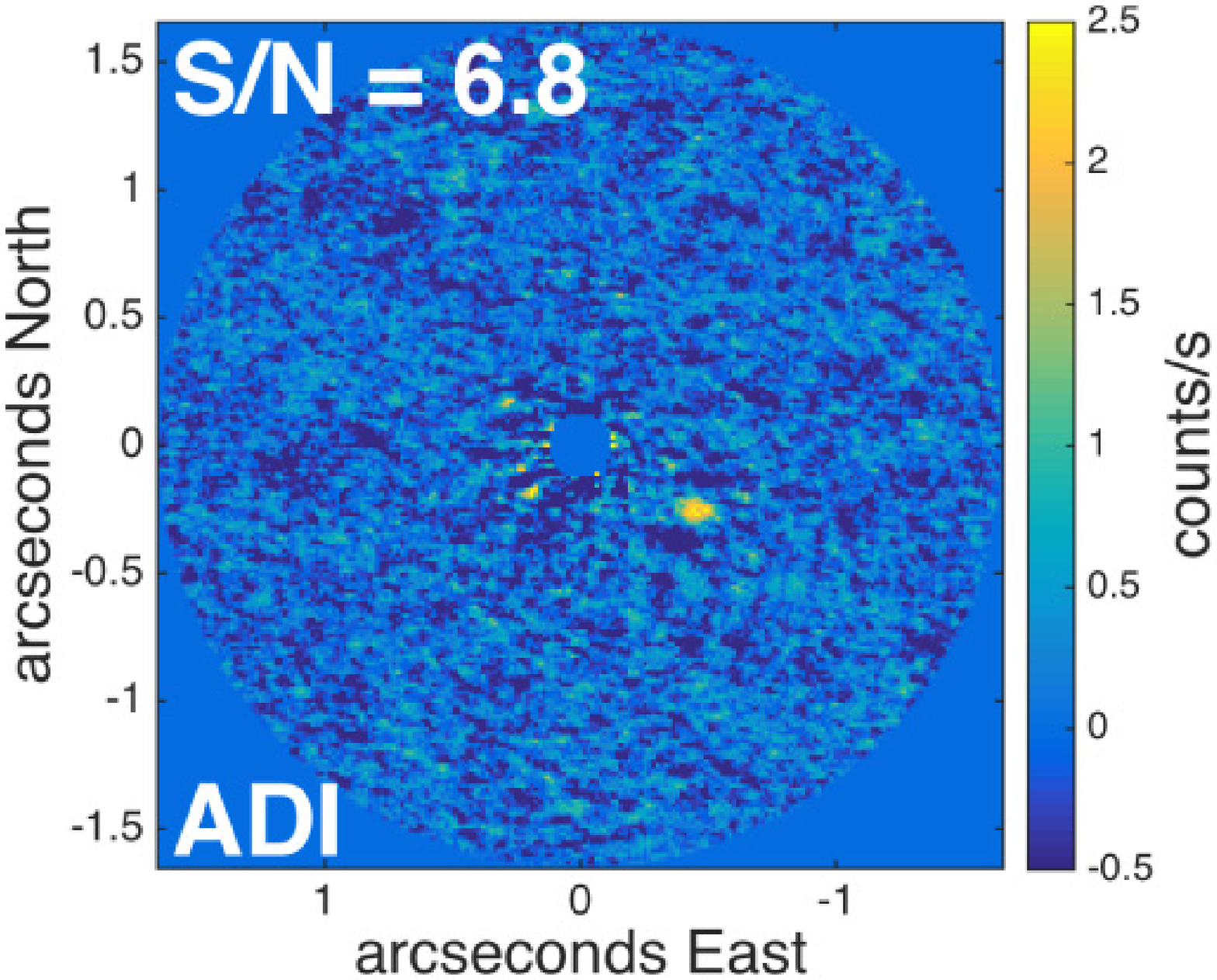}}
\subfloat[]{\label{fig:BDIfakeclose}\includegraphics[width=0.49\textwidth]{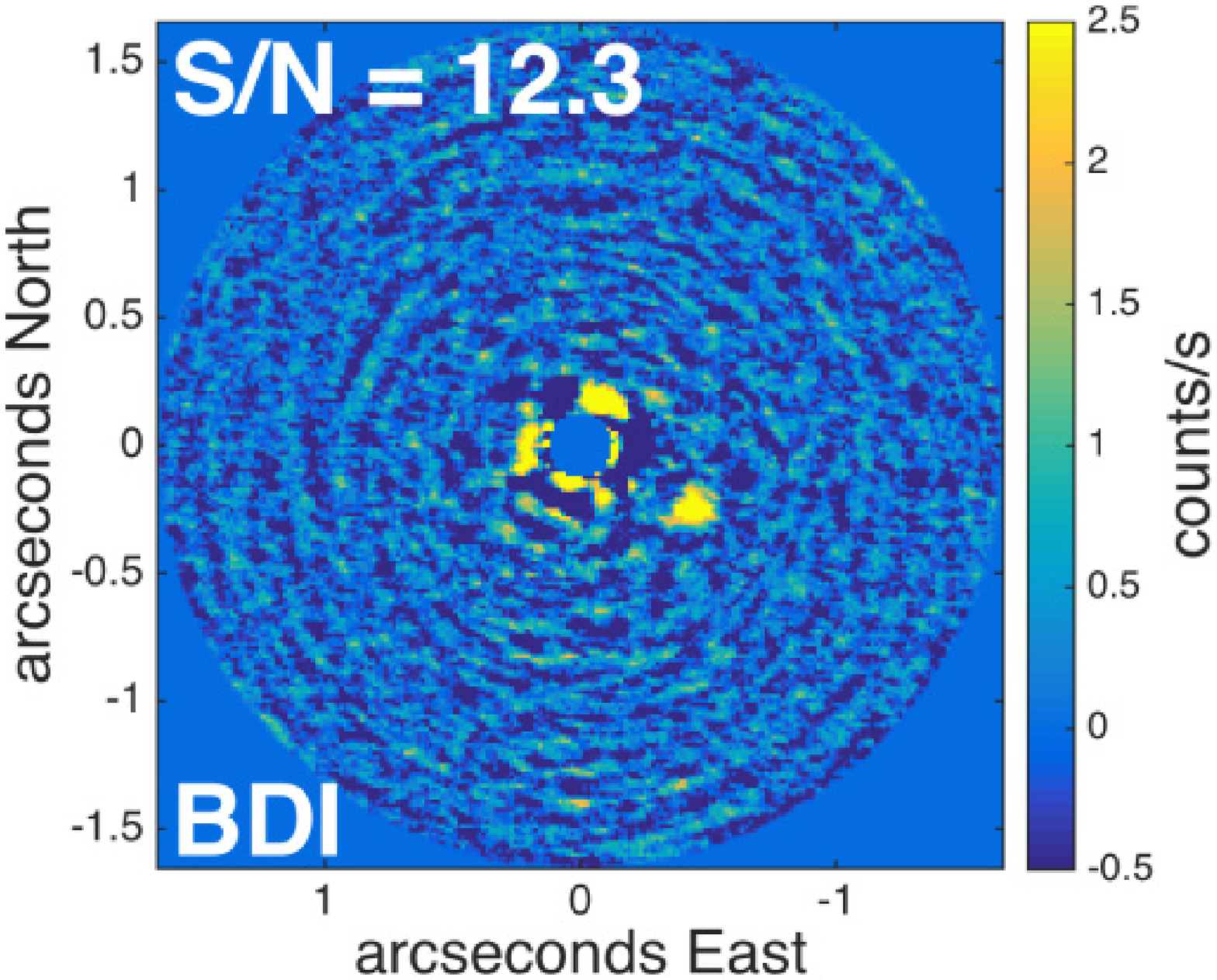}} 
%\subfloat[]{\label{fig:percent}\includegraphics[width=0.33\textwidth]{BDIpercentA.eps}}
\caption{Examples of inserted and recovered artificial companions. (a-b): ADI- and BDI-recovered $\Delta m = 6$ planets at $r$ = 1\fasec 25 and $\theta = 95$\degrees. (c-d): The same, but at $r$ = 0\fasec 5 and $\theta = 240$\degrees. In both cases, the S/Ns of the BDI-recovered planets are higher than in the ADI case. It is also evident that the BDI images contain significant residuals very close to the star, unlike the ADI images. This indicates that the binary PSFs were not perfectly identical. Nonetheless, BDI still outperforms ADI at small separations because ADI significantly attenuates companion flux.}
\label{fig:fake}
\end{figure*}

To fully test the capabilities of BDI, we computed the overall contrast limits achieved by ADI and BDI. We accomplished this using a Monte Carlo approach similar to the one employed in \cite{me4796} whereby a single artificial planet with a random contrast ($\Delta m$) at a random separation ($r$) and position angle ($\theta$) is repeatedly inserted into the raw data and recovered using both ADI and BDI. We inserted and recovered a total of 10,000 artificial planets, which were flux-scaled replicas of the Star A PSF. The allowed parameter ranges were $r \in $ [0\fasec 15, 1\fasec 55], $\theta \in $ [0, 2$\pi$), $\Delta m \in $ [5, 9]. Fig. \ref{fig:fake} shows examples of artificial planets recovered by both BDI and ADI.

\begin{figure*}[t!]
\centering
\subfloat[]{\label{fig:rawcontrast}\includegraphics[width=0.49\textwidth]{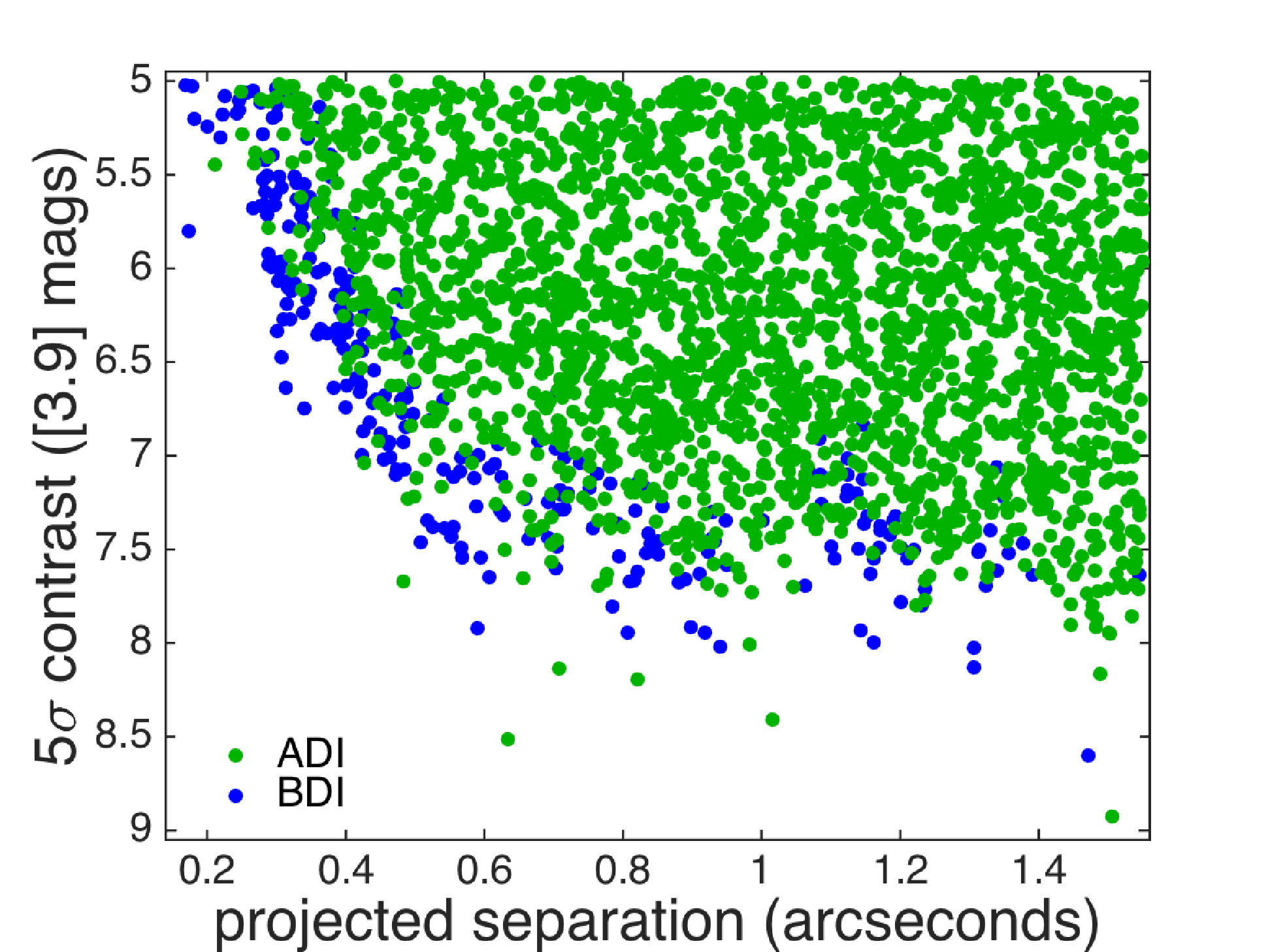}}
\subfloat[]{\label{fig:winning}\includegraphics[width=0.49\textwidth]{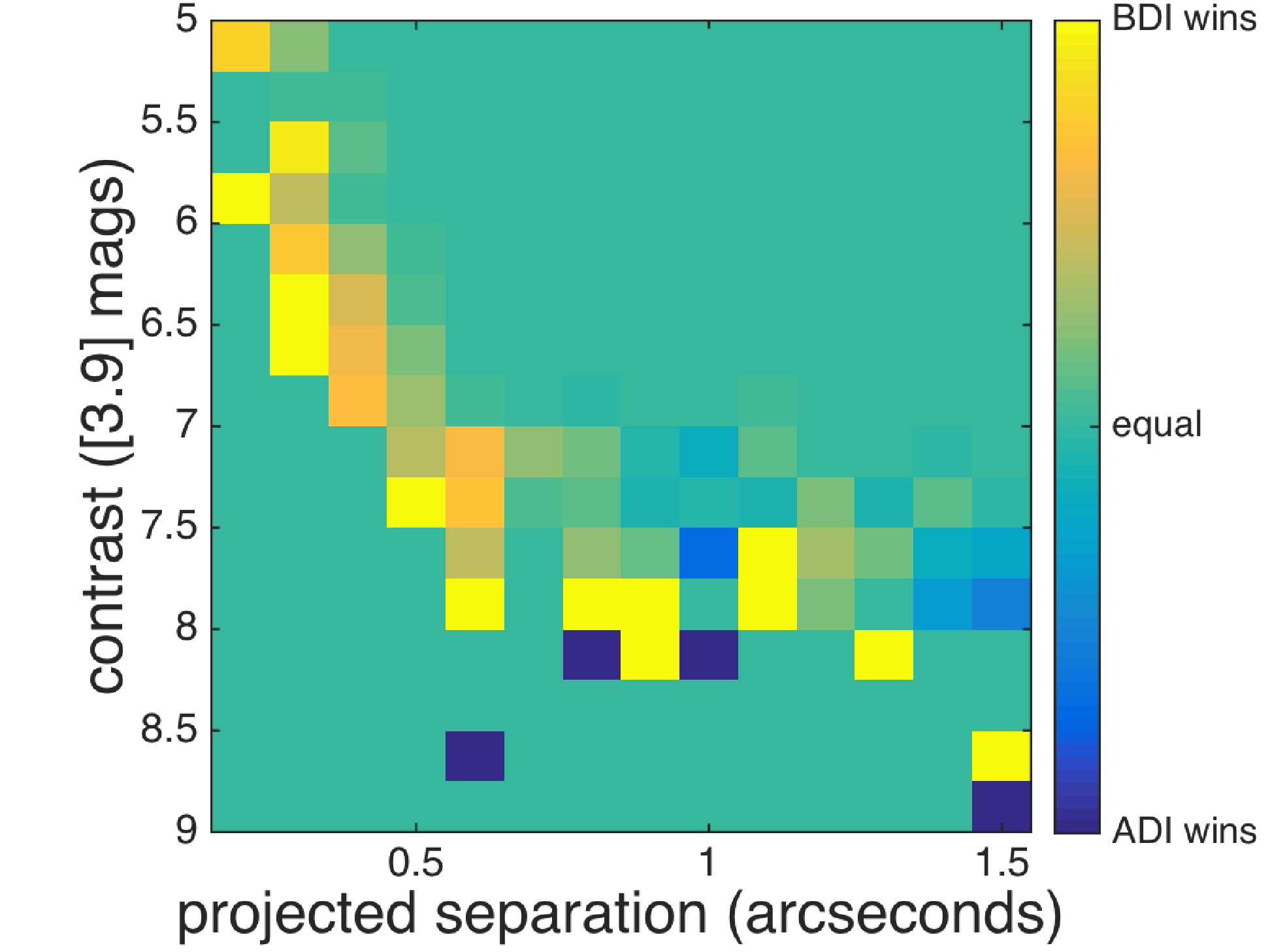}}
\caption{(a): 5$\sigma$ contrast for artificial planets recovered by ADI and BDI. The green points denote planets that were detected at S/N $\geq$ 5 by ADI, while the blue points (most of which are hidden by the green points) denote those that were detected by BDI. BDI generally detects planets closer to the star than ADI and is \about 0.5 mags better in terms of contrast within \about 1\asec. (b): Binned contrast map showing the frequency of successful detections by BDI compared to ADI. Yellow squares contain more S/N $\geq$ 5 detections by BDI than ADI, and vice versa for the blue squares. Green squares contain equal numbers of detections by both methods. BDI tends to perform better closer to the star than ADI.}
\label{fig:contrast}
\end{figure*}

We computed the S/Ns of the recovered artificial planets using the updated definition of S/N from \cite{mawetsn}, which is appropriate for speckle-dominated regions close to the star. Fig. \ref{fig:rawcontrast} shows the separation and contrast values of planets that were detected with S/N $\geq$ 5 by ADI and BDI. BDI generally detects planets closer to the star than ADI, being \about 0.5 mags better in terms of contrast within \about 1\asec. Another way of showing this is using a binned contrast map, wherein we counted the number of successful detections (S/N $\geq$ 5) by both ADI and BDI in 0\fasec 1 $\times$ 0.25 mag boxes. Fig. \ref{fig:winning} shows that BDI tends to detect a higher fraction of planets closer to the star than ADI. In terms of planet masses, using the COND mass-luminosity evolutionary tracks \citep{baraffe} for a 130 Myr old star, a half magnitude improvement translates into \about 1 Jupiter mass improvement in sensitivity for BDI over ADI.

%As in \cite{me4796}, we constructed contrast ``maps" for both the ADI-recovered and BDI-recovered planets. We binned (mean) the S/N values in 0\fasec 1 $\times$ 0.25 mag boxes. Fig. \ref{fig:contrastmap} shows the contrast maps for ADI and BDI. 

\section{Discussion}
\subsection{Benefits of BDI}
We have shown with 3.9 \microns ~MagAO/Clio-2 data that when two stars are imaged simultaneously at high Strehl ratio within the isoplanatic patch, BDI can be used to potentially probe smaller planet masses at closer separations than in the ADI case. In addition, BDI offers two unique technical advantages. First, PA rotation is no longer a concern because each component is used as the other's PSF reference. This is a major improvement with regard to observing efficiency and sky coverage and offers the observer much more flexibility in planning an observational sequence over the course of a night (although non-isoplanatism may become a concern at high airmass). Second, because BDI allows us to search both stars for companions simultaneously, binaries can be studied 2$\times$ (4$\times$) more efficiently than with ADI (classical PSF subtraction).  

%This is because in typical ADI surveys, single stars are imaged, whereas with BDI two stars are imaged at the same time. This means that planets can be detected around \textit{both} stars simultaneously. Therefore, for a typical ADI survey of 100 stars, BDI would only need 50 binaries to reach the same sample size. This also means that fewer telescope hours would be needed to achieve the same survey goals, an important consideration for time allocation committees on future exoplanet imaging instruments. 

%For the case of a telescope in space, BDI is again 2$\times$ more efficient than PSF subtraction obtained by``rolling" the telescope. BDI is 4$\times$ more efficient than classical PSF subtraction because two targets must be observed sequentially to search for planets around one star. 

%ADI requires that stars typically be imaged through transit to maximize sky rotation over short periods of time. However, observing through transit is an optimistic case and is not always achieved; PA rotations are thus often reported to be $< 30$\degrees ~in exoplanet imaging studies, limiting the detectability and S/N of planets at close separations. For BDI, the only concern is the visibility of the target (i.e., airmass), which offers the observer much more flexibility in planning an observational sequence over the course of a night (although non-isoplanatism may become a concern at high airmass).

%\subsection{The Science Case for BDI}
From an astrophysical perspective, BDI offers the unique advantage of targeting stars (binaries) that are not usually preferred in large direct imaging surveys.\footnote{\cite{thalmannspots} are conducting a direct imaging search for planets in spectroscopic binaries. However, the detectable planets in that survey are ones that orbit \textit{both} stellar components (P-type orbits). Such planets are presumably distinct from planets that orbit single stars.} The S-type planets (planets orbiting a single component of the binary) that BDI would be sensitive to should form and remain mostly unhindered by distant stellar companions \citep{binaryplanetfrequency}. Furthermore, we know that \about half of ``hot Jupiter" systems have stellar companions, and these companions seem to have no effect on the primary star-planet orbital alignment \citep{friendsofhotjupiters2}. It is thus important to search for and detect planets in binaries and test theoretical predictions for the formation and stability of such planets \citep{binarystability,jangcondellbinaries}. From a more general perspective, because \about half of all stars reside in multiple systems \citep{binaryfrequency}, it is important to search for planets around the other half of stars that are typically excluded \citep{thalmannspots}. 

As a first attempt at probing planet formation and evolution in wide binaries, we are conducting a BDI survey for young exoplanets in visual binaries using MagAO/Clio-2. To minimize noise (e.g., Section \ref{sec:theory}), we have selected binaries that have comparable brightnesses in the NIR ($\Delta m \lesssim$ 2).\footnote{This is not a strict requirement. As described in Section \ref{sec:theory}, a larger brightness ratio means that the noise will be smaller when the brighter star is subtracted from the fainter one, but it will also be larger in the opposite case. This essentially reduces the efficiency of the survey because sensitivity is better for one star and worse for the other.} Their projected separations are between \about 2-10\asec ~such that they lie within the isoplanatic patch at 4 \microns, their PSFs do not overlap, and they fit on the Clio-2 detector (size 9\asec$\times$15\asec). The primary component of each binary has $V \lesssim 15$ so that the AO system can lock. The binaries are young ($\lesssim$ 500 Myr) and close to Earth ($\lesssim$ 150 pc). Based on these constraints, the survey sample consists of \about 140 binaries (280 total stars) in young associations in the southern hemisphere. 

%A BDI survey for planets in visual binaries would improve our understanding of the demographics of giants planets orbiting close to their host stars. It would also help answer the following fundamental questions: (1) What are the atmospheric properties of directly imaged planets? (2) What is the frequency and distribution of young, gas giant planets in young visual binaries? (3) How do the planet properties depend on the binary properties (stellar masses, semimajor axes, eccentricities)? In particular, what effects do wide binaries have on planet formation \citep{binarystability}? 

%\subsection{The MagAO BDI Survey}
%\subsubsection{Ground-based with Adaptive Optics}

\subsubsection{Potential Space-based BDI Survey}
BDI is not limited to ground-based telescopes with AO. The JWST NIRCam instrument will offer space-based NIR imaging with a large 2.2$\arcmin$ FOV. This opens up the possibility of imaging very wide binaries--or even single stars that happen to be close to other field stars--greatly expanding the potential survey size. Furthermore the similarity of two simultaneously-imaged PSFs should be limited by the (likely small) variations in the detector (like distortion) rather than the atmosphere/AO correction. 

%As discussed in Section \ref{sec:theory}, the noise should be at worst \about a few $\times$ higher than in coronagraphic imaging, with the additional benefit of a better IWA and 4$\times$ better observational efficiency. 

The number of possible survey stars would be restricted by the saturation of the NIRcam detector and the amount of PSF overlap. Using the JWST NIRcam PSF simulator (\textit{webbpsf}\footnote{\url{http://www.stsci.edu/jwst/software/webbpsf}}), in the absence of noise 98$\%$ of the PSF flux is contained with \about 5\asec. Therefore stars with separations between \about 5-60\asec ~could be observed. Assuming bright stars ($V < 6$) saturate the detector, there are conceivably \about several hundred young visual binaries and young single stars (located close to other field stars) that could constitute a deep and powerful exoplanet imaging survey. 

\subsection{Limitations}
\label{sec:limitations}
%\begin{figure}[t!]
%\centering
%\includegraphics[width=0.49\textwidth]{snplots3.eps}
%\subfloat[]{\label{fig:contrastcurve}\includegraphics[width=0.49\textwidth]{contrastcurve.eps}}
%\includegraphics[width=0.49\textwidth]{BDIpercentA.eps}
%\caption{The same as Fig. \ref{fig:percent}, except for ADI planets that were recovered with a rotation requirement of 1 FWHM. Uncertainties are the errors on the mean in each bin. While this helps to preserve companion flux, it does not improve overall recovered S/N and thus is still generally inferior to BDI.}
%\label{fig:fwhm}
%\end{figure}

Here we discuss possible limitations of our study and the BDI technique. First, we tested the ADI/BDI comparison as a function of achieved sky rotation, since ADI works better with larger rotations. We divided the data in half (94 images in the first half, 95 in the second half). The PA changed by a total of \about 10\degrees ~in each half, and the total integration time for each half was \about 15 minutes. We injected and recovered artificial planets (as in Section \ref{sec:fake}) and compared the recovered S/Ns to the S/Ns achieved using the entire data set (full rotation and integration). We found that while the S/Ns of ADI-recovered planets increased with more rotation, the S/Ns of BDI-recovered planets increased by similar (or larger) amounts. This is because BDI is essentially integration-limited, rather than rotation-limited, so doubling the integration time invariably increases the contrast. Furthermore, increased sky rotation also benefits BDI by azimuthally-smearing out residuals \citep{adi}.

%The only case where ADI might have a chance to perform as well as BDI would be in very short integrations over which the PA changes significantly. Then ADI-recovered planets will not be hampered by self-subtraction, but will rather be limited by integration (and PSF stability; see Section \ref{sec:theory}). However, most direct imaging searches try to maximize both PA rotation and integration, in which case BDI would again be expected to win out.

Second, we tested the use of a rotation threshold in the ADI data reduction. This would be expected to reduce companion self-subtraction and thereby possibly increase companion S/N; however, this would also mean that $\sigma_k$, the noise from speckles in PSF subtraction, would be larger due to the increased dissimilarity between the target and reference PSFs. Therefore it is not obvious that the overall S/N should increase. To test this, we inserted 1000 artificial planets (as in Section \ref{sec:fake}) and recovered them with BDI and ADI. However, in the ADI case, for each image we required that the PSF be constructed from images that had rotated by at least 1 full-width half-maximum (FWHM) at the companion's location. We found that the inclusion of a rotation threshold did not significantly increase the ADI-recovered companion S/N.

Third, we tested the combination of ADI and BDI in the data reduction to potentially increase detection sensitivity. Using our MagAO data, we tried running BDI first, then ADI, and vice versa (with ADI on each star). We also tested making all images (Star A and Star B) available for PSF reconstruction, as well as the same but with a rotation threshold on the Star A images. Overall, achieved contrast was not improved compared to BDI alone. However, in other cases (different brightness ratios and/or binary separations), an ADI+BDI combination may prove more useful than either technique alone.

A limitation of BDI (that ADI is not affected by) is the possibility of companion flux attenuation by \textit{planets around the other star}. For example, if both stars have planets at similar separations and position angles, then the flux of each will be attenuated by the other. However, even if two planets have identical brightnesses, the final fluxes will only be partially attenuated because each planet's flux is smeared out by PA rotation during the observations. Furthermore, assuming no preference for planet locations in an image, the chances of two planets being at the same locations around two stars are \about $2\%$ at 0\fasec 15 and much smaller farther out. 

Finally, it should be noted that coronagraphs, which have become increasingly utilized in exoplanet imaging efforts, cannot be used in BDI. Specifically, coronagraphs in the image plane cannot be used. This limits the sensitivity of any BDI survey. However, it would be possible to use the apodizing phase plate (APP, \citealt{app,newapp,newapp2}), a pupil-plane ``coronagraph" that offers improved contrast close to the star. Because the device sits in the instrument pupil plane, it affects all stars imaged on the detector. MagAO/Clio-2 already has several APPs installed, which can be used to further increase planet detection sensitivity with BDI. 

%Finally, one could make the case that our proof of concept data was biased towards favoring BDI because the images were essentially sky background-limited up to \about 0\fasec 3. This results from the star being faint ($V = 9.65$). If we had observed a brighter target, then the images may have been contrast-limited at wider separations. In the contrast-limited regime, speckle noise dominates, so ADI may have fared better than BDI despite the companion self-subtraction. 

% sky limited vs. contrast limited

\section{Summary}
We have shown binary stars should be considered viable future targets for large direct imaging surveys because of the potential for improved PSF subtraction close to the stars using BDI. This is important because planets are thought to be rare at wide separations. Based on the results of our proof of concept BDI data, we are conducting a new direct imaging search for exoplanets in young visual binaries with MagAO/Clio-2. This survey should be able to detect fainter planets closer to their host stars and will tell us about exoplanet demographics in stellar populations that have never been probed before. BDI on a space-based telescope like JWST/NIRcam would not be limited by isoplanatism effects and would therefore be an even more powerful technique for imaging and discovering exoplanets.

\acknowledgments
We thank the anonymous referee for helpful comments and suggestions. EEM acknowledges support from NSF grant AST-1313029. J.R.M was supported under contract with the California Institute of Technology (Caltech) funded by NASA through the Sagan Fellowship Program.

%\clearpage

\bibliographystyle{apj}
\bibliography{ms}

\end{document}